\documentclass[10pt]{article}
\usepackage{fullpage}
\usepackage{amsmath}
\usepackage{amssymb}
\usepackage{amsfonts}
\usepackage{comment}
\usepackage{color}
\usepackage{cite}
\usepackage{subcaption}
\usepackage{graphicx}
\usepackage{enumitem}

\def\blue{\textcolor{black}}

\def\##1{{\bf #1}}
\def\=#1{\underline{\underline{#1}}}

\def\+
#1{\underline{\bf #1}}
\def\*#1{\underline{\underline{\bf #1}}}

\def\r#1{(\ref{#1})}
\def\l#1{\label{#1}}

\def\le{\left(}
\def\ri{\right)}
\def\les{\left[}
\def\ris{\right]}
\def\lec{\left\{}
\def\ric{\right\}}
\def\lek{[{\kern 0.1em}}
\def\rik{{\kern 0.1em}]}

\def\.{\mbox{ \tiny{$^\bullet$} }}

\def\eps{\varepsilon}

\def\epso{\eps_{\scriptscriptstyle 0}}

\def\muo{\mu_{\scriptscriptstyle 0}}

\def\ko{k_{\scriptscriptstyle 0}}

\def\ux{\hat{\#u}_{\rm x}}
\def\uy{\hat{\#u}_{\rm y}}
\def\uz{\hat{\#u}_{\rm z}}

\begin{document}

\begin{center}

\LARGE{ {\bf Depolarization dyadics for truncated spheres, spheroids, and ellipsoids
}}
\end{center}
\begin{center}
\vspace{10mm} \large
 
 {Tom G. Mackay}\footnote{E--mail: T.Mackay@ed.ac.uk.}\\
{\em School of Mathematics and
   Maxwell Institute for Mathematical Sciences\\
University of Edinburgh, Edinburgh EH9 3FD, UK}\\
and\\
 {\em NanoMM~---~Nanoengineered Metamaterials Group\\ Department of Engineering Science and Mechanics\\
The Pennsylvania State University, University Park, PA 16802--6812,
USA}
 \vspace{3mm}\\
 {Akhlesh  Lakhtakia}\\
 {\em NanoMM~---~Nanoengineered Metamaterials Group\\ Department of Engineering Science and Mechanics\\
The Pennsylvania State University, University Park, PA 16802--6812, USA}

\normalsize

\end{center}

\begin{center}
\vspace{5mm} {\bf Abstract}
\end{center}

Depolarization dyadics play a central role in theoretical studies involving scattering from small particles and homogenization of particulate composite materials. 
 Closed-form expressions for depolarization dyadics have been developed for
  truncated spheres and truncated spheroids, and the formalism has been extended to truncated ellipsoids; the evaluation of depolarization dyadics for this latter case requires numerical integration. The H\"{o}lder continuity condition has been exploited to fix the   origin of the coordinate system for the evaluation of depolarization dyadics.
 These  results will enable theoretical studies involving scattering from small particles and  homogenization
 of  particulate composite materials to accommodate particles with a much wider range of shapes than was the case hitherto.

 \vspace{5mm}
\vspace{5mm}

\section{Introduction}

 The calculation of the electric field due to a specified source current density is a fundamental problem in  electromagnetics \cite{Chen}.
 The usual approach involves integration of a suitable product of a dyadic Green function and the source
 current density over the source region (i.e., the region occupied by the source current density) \cite{Tai,Faryad}.
  This process is particularly challenging if the electric field is sought in the source region, since  the singularity of dyadic Green function must be considered then \cite{VanBladelBook}. 
  The integrated singularity of the dyadic Green function~---~known as the \emph{depolarization dyadic}~---~is a
  central pillar in scattering and homogenization theories \cite{EAB,MAEH}.
  
The depolarization dyadic is expressible in terms of a surface integral. The evaluation of this integral is crucially dependent upon the shape of the surface. For relatively simple shapes, such as spherical \cite{VanBladelBook}, spheroidal \cite{M97,Moroz}, ellipsoidal \cite{Osborn, Stoner,MW97,W98}, cylindrical \cite{Lee,WM02}, cubical \cite{Lee}, and polyhedral \cite{L98}, closed-form expressions for depolarization dyadics have been developed, but for more complex shapes numerical methods must be used for their evaluation. 

In this communication, closed-form expressions are developed for truncated spheres and spheroids, and these are illustrated numerically. The formalism is extended to truncated ellipsoids.
Hereafter, $\epso$, $\muo$, and $\ko = \omega \sqrt{\epso \muo}$ are the free-space permittivity, permeability, and wavenumber, respectively, with $\omega$ being the angular frequency; an $\exp(-i \omega t)$ time-dependence is implicit; and 
$\=I=   \ux\ux+ \uy\uy+ \uz\uz $ is the identity dyadic, with $\ux$, $\uy$, and $\uz$ being unit vectors aligned with the Cartesian coordinate axes .

\section{Depolarization dyadics}

Suppose that 
 a time-harmonic source current density $\#J(\#r)$ exists inside a finite  region $V_s$ which is  bounded by a closed surface.
The unbounded region outside $V_s$ is vacuous.
The time-harmonic electric field  both outside and inside $V_s$ is given as \cite{Chen}
\begin{equation} \l{E}
\#E (\#r) = i \omega \muo \int_{V_s} \=G(\#r, \#r_s) \. \#J(\#r_s) \, d^3 \#r_s, 
\end{equation}
wherein
\begin{equation} \l{DGF}
\=G(\#r, \#r_s) = \le \=I + \frac{\nabla \nabla}{\ko^2} \ri \frac{\exp \le i \ko |\#r - \#r_s |\ri}{4 \pi | \#r - \#r_s |} 
\end{equation}
is  the free-space dyadic Green function  \cite{Faryad}.

For field points outside the source region, i.e.,  $\#r \notin V_s$,
the integral on the right side of Eq.~\r{E} is well behaved and its evaluation delivers $\#E(\#r)$ as an analytic function.
For field points inside the source region, i.e., $\#r \in V_s$, the evaluation of the integral on the right side of Eq.~\r{E}  requires care  
because
 the dyadic Green function is singular at $\#r = \#r_s$ \cite{VanBladelBook}. 
 In particular, the double derivative  in the free-space dyadic Green function \r{DGF} gives rise to a $|\#r - \#r_s|^{-3}$ term. This is not integrable unless
  the source current density $\#J(\#r)$  
 satisfies the following H\"{o}lder continuity condition: 
 there exist
  three positive constants $a_1$, $a_2$, and
  $a_3$  such that \cite{Wang}
  \begin{equation} \l{Holder}
  | \#J(\#r) - \#J(\#r_s)| \leq a_1 | \#r - \#r_s |^{a_2}
  \end{equation}
 for all $\#r_s$ satisfying $|\#r - \#r_s| \leq a_3$.

\begin{figure}[!htb]
\begin{subfigure}{.3\textwidth}
\centering
\includegraphics[width=5.5cm]{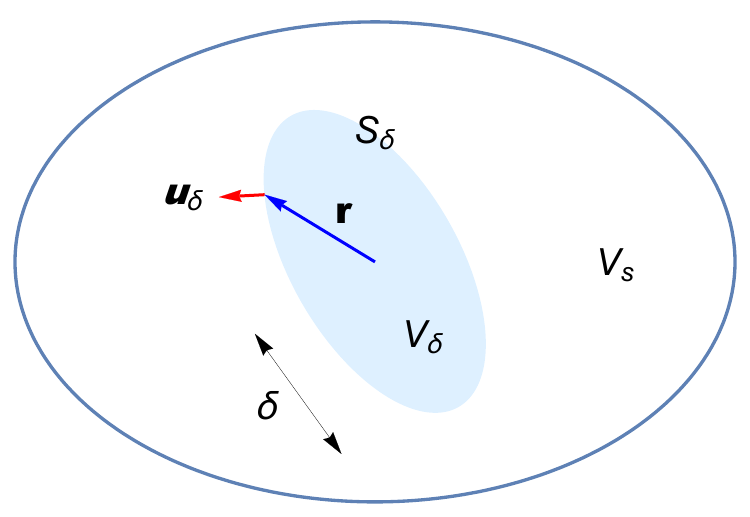} 
  \vspace{0mm}  \hfill
 \caption{\label{Fig1a} Exclusion region $V_\delta$ inside source region $V_s$.
   }
   \end{subfigure}
   \begin{subfigure}{.3\textwidth}
\centering
\includegraphics[width=5.5cm]{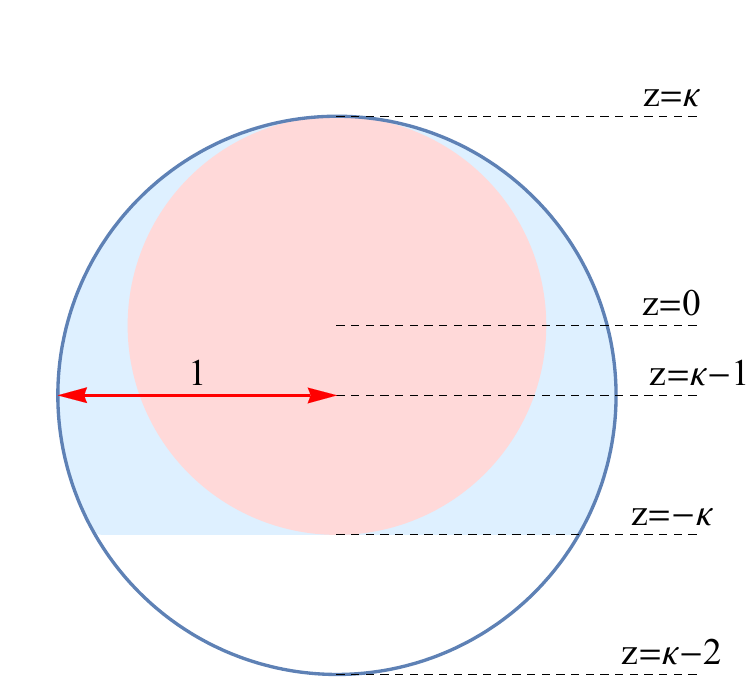} 
  \vspace{0mm}  \hfill
 \caption{\label{Fig1b} Truncated unit sphere with largest inscribed sphere of radius  $\kappa\in(0,1)$.
   }
\end{subfigure}
\begin{subfigure}{.3\textwidth}
\centering
\includegraphics[width=5.5cm]{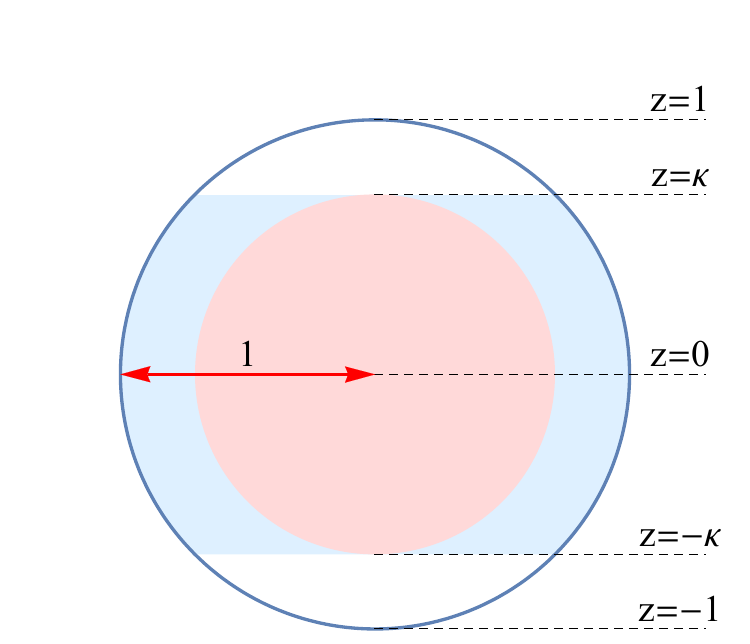} 
  \vspace{0mm}  \hfill
 \caption{\label{Fig1c} Double-truncated unit sphere with largest inscribed sphere of radius $\kappa\in(0,1)$.
   }
\end{subfigure}
\begin{subfigure}{.5\textwidth}
\centering
\vspace{20mm}
\includegraphics[width=5.5cm]{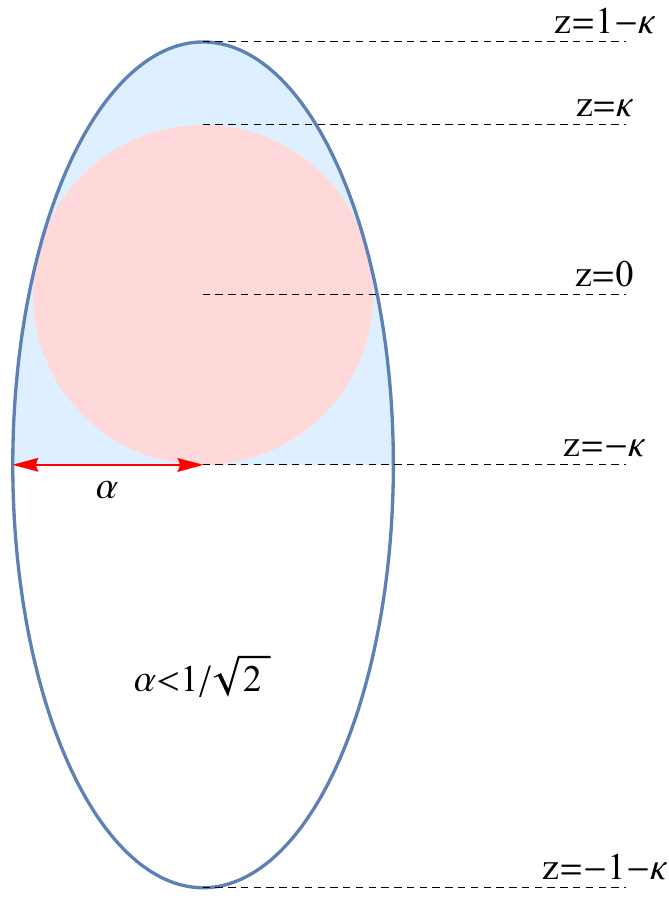} 
\includegraphics[width=5.5cm]{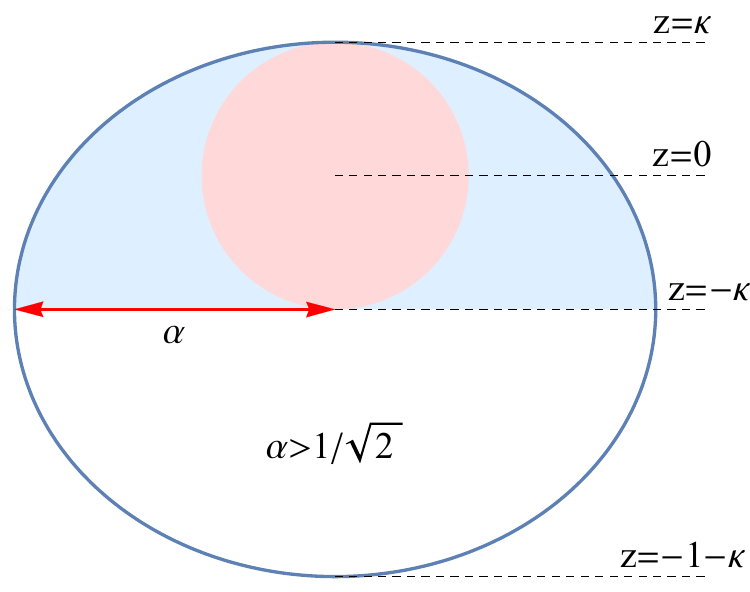} 
  \vspace{0mm}  \hfill
 \caption{\label{Fig1d} Hemispheroid with largest inscribed sphere of radius  $\kappa\in(0,1)$ for 
  $\alpha < 1/\sqrt{2}$ (top) and
 $\alpha > 1/ \sqrt{2}$ (bottom).
   }
\end{subfigure}
\hspace{10mm}
\begin{subfigure}{.5\textwidth}
\centering
\vspace{20mm}
\includegraphics[width=5.5cm]{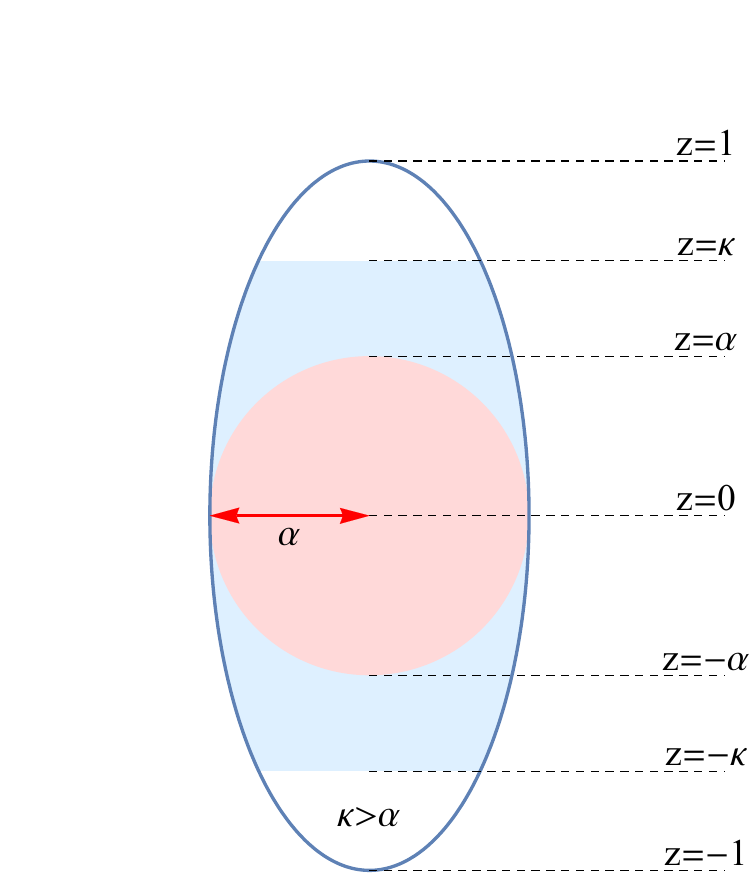} 
\hspace{10mm}
\includegraphics[width=5.5cm]{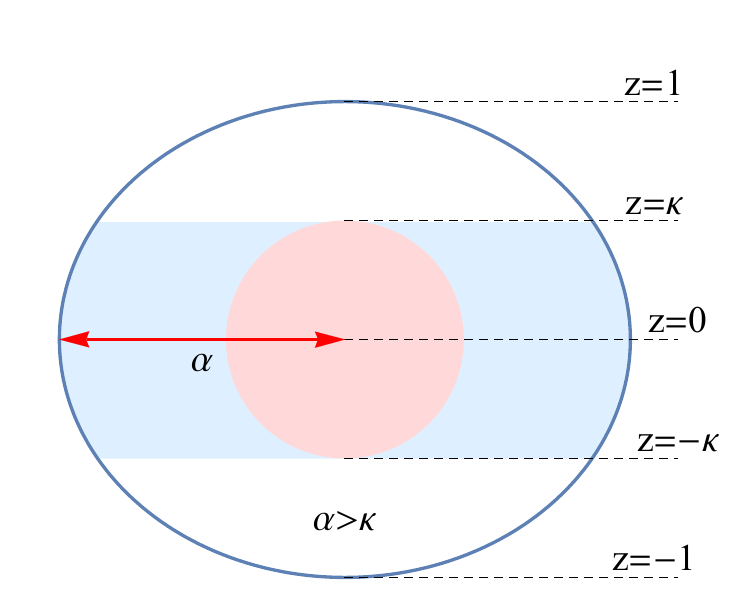} 
  \vspace{0mm}  \hfill
 \caption{\label{Fig1e} Double-truncated spheroid with largest inscribed sphere of radius $\alpha$  for 
  $\alpha < \kappa$ (top) and of radius $\kappa$  for
 $\alpha > \kappa$ (bottom).
   }
\end{subfigure}
 \caption{\label{Fig1} Schematic representations of exclusion regions.
   }
\end{figure}

 Provided that the  H\"{o}lder continuity condition \r{Holder} is satisfied, the electric
field inside the source region $V_s$ may be expressed as \cite{Yaghjian}
 \begin{equation} \l{Esource}
\#E (\#r) = i \omega \muo \lim_{\delta \to 0} \int_{V_s - V_\delta} \=G(\#r, \#r_s) \. \#J(\#r_s) \, d^3 \#r_s + \frac{1}{i \omega \epso} \=L\. \#J(\#r), \qquad \#r \in V_s.
\end{equation}
Herein \blue{$V_\delta \subset V_s$ is an exclusion region of linear dimensions} quantifiable through
a small length $\delta$, and surface $S_\delta$, that contains the singular point $\#r = \#r_s$; a schematic representation is displayed in Fig.~\ref{Fig1a}.
The depolarization dyadic
 \begin{equation} \l{depol}
 \=L = \frac{1}{4 \pi} \int_{S_\delta} \frac{\#u_\delta \, \#r}{|\#r|^3} \, d^2 \#r,
 \end{equation}
 with $\#u_\delta$ being the unit outward normal vector to $S_\delta$.
\blue{The shape
of $V_\delta$ should be such that 
$\#u_\delta$
is unambiguously identified at every point on $S_\delta$, but
edges
on $S_\delta$
 can be rounded off slightly to overcome this restriction, if necessary.}

The question arises: Where inside the exclusion region $V_\delta$ should the coordinate origin be taken for the integration that delivers the depolarization dyadic in Eq.~\r{depol}? In the case of the spherical exclusion region, there is no problem because 
 $\=L = (1/3) \=I$ regardless of where the coordinate origin is located inside $V_\delta$  \cite{VanBladelBook}. However, 
 $\=L$ 
generally
depends on the choice of coordinate origin for less symmetric exclusion regions  \cite{Yaghjian,Sihvola}.
The origin of the coordinate system must therefore be taken as the center of the largest sphere that can be inscribed inside the exclusion region $V_\delta$, so that  the   H\"{o}lder continuity condition \r{Holder} holds over the largest portion of $V_\delta$.

In the following sections, depolarization dyadics are calculated for exclusions regions shaped as truncated spheres, spheroids, and ellipsoids.
 For a truncated ellipsoid whose principal axes are aligned with the axes of the Cartesian coordinate system, the depolarization
dyadic has the diagonal form
\begin{equation}
 \=L  =   L_x\, \ux\ux+ L_y\,\uy\uy+ L_z\,\uz\uz ;
\end{equation}
with  $L_x= L_y \equiv L$ for  truncated spheres and truncated spheroids.
Furthermore,
the trace of the depolarization dyadic
\begin{equation}
 \mbox{trace} \lec \=L \ric = \frac{1}{4 \pi} \int_{S_\delta} \frac{\#u_\delta \. \#r}{|\#r|^3} \, d^2 \#r
 \end{equation}
 represents
  the normalized solid angle of the surface $S_\delta$ as viewed from $\#r = \#0$ \cite{Yaghjian}. Hence, $ \mbox{trace} \lec \=L \ric  = 1$. Accordingly, in the following presentation of closed-form expressions for depolarization dyadics for truncated spheres and spheroids, there is no need to  provide explicit expressions for $L_z$ because $L_z = 1- 2L$.

 \section{Spherical geometry}
 
\subsection{Truncated sphere} \l{Truncated_sphere_sec}

Suppose that the unit sphere centered at $x=y=0, z= \kappa -1 $, i.e.,
\begin{equation}
x^2 + y^2 + \le z - \kappa +1 \ri^2  \leq  1,
\end{equation}
 is bifurcated by the plane $z = -\kappa$, where  $0<\kappa<1$, as schematically illustrated in Fig.~\ref{Fig1b}. 
 The exclusion region $V_\delta$ is the upper part of the sphere
bounded below by  the plane $z = -\kappa$, and
the 
 largest inscribed sphere is specified by
\begin{equation} \l{largest_sphere}
x^2 + y^2 + z^2 \leq \kappa^2.
\end{equation}
After integrating over both the curved and the flat parts of $S_\delta$,
we found  from  Eq.~(\ref{depol}) that $\=L  =   L \left( \ux\ux+  \uy\uy\right)+ L_z\,\uz\uz$ with
\begin{equation}
L =  \frac{ \kappa  }{6 \le 1 - \kappa\ri^3 \tau}\,\lec 6 - \tau \le 3  - 3 \kappa  + \kappa^2 \ri - \kappa \les 11 - 3 \kappa \le 3- \kappa 
\ri\ris \ric,
\end{equation}
where
\begin{equation}
\tau = \sqrt{\le 4 - 3 \kappa \ri \kappa};
\end{equation}
and confirmed that $L_z = 1- 2L$.
Observe that $L \to 0$ and $L_z \to 1$ in the limit $\kappa \to 0 $, while
$L \to 1/3$ and $L_z \to 1/3$ in the limit $\kappa \to 1 $, in agreement with well-known results \cite{VanBladelBook}. 

\begin{figure}[!htb]
\centering
\includegraphics[width=8cm]{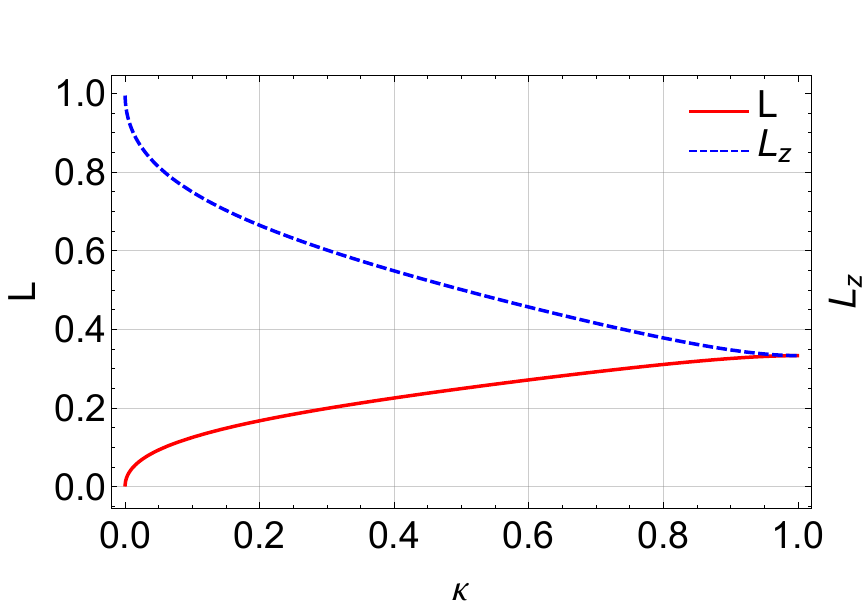} 
  \vspace{0mm}  \hfill
 \caption{\label{Fig2} $L$ and $L_z$ plotted against  $\kappa\in(0,1)$ for the truncated unit sphere.
   }
\end{figure}

Plots of the depolarization factors $L$ and $L_z$ versus $\kappa$ are provided in Fig.~\ref{Fig2}. As $\kappa$ increases,
the value of $L$ increases monotonically  whereas the value of $L_z$ decreases monotonically. For all $\kappa < 1$, $L_z > L$.
The case of the hemisphere, i.e., $\kappa = 1/2$, 
for which $L = (19/6\sqrt{5})- 7/6 = 0.24951$ and $L_z = \le 50 - 19 \sqrt{5} \ri /15 = 0.500981 $ 
 is notable: $L_z$ is approximately~--~but not exactly~--~equal to $ 2 L$.

\subsection{Double-truncated sphere}

Suppose that the unit sphere centered at the origin of the coordinate system, i.e.,
\begin{equation}
x^2 + y^2 + z^2 \leq 1,
\end{equation}
 is symmetrically trifurcated   by the planes $z = -\kappa$ and   plane $z=\kappa$, where  $0<\kappa<1$, per 
  the schematic representation in Fig.~\ref{Fig1c}. The exclusion region $V_\delta$ is
   the double-truncated sphere bounded by the planes $z = \pm \kappa$.
  
  The depolarization dyadic
 for this $V_\delta$   again
 is of the form $\=L  =   L \left( \ux\ux+  \uy\uy\right)+ L_z\,\uz\uz$.
  With the 
 largest inscribed sphere   specified by Eq.~\r{largest_sphere},
we determined
\begin{equation}
L = \frac{ \le 3- \kappa^2 \ri \kappa}{6} 
\end{equation}
from  Eq.~(\ref{depol})
and verified that   $L_z=1-2L$.

\begin{figure}[!htb]
\centering
\includegraphics[width=8cm]{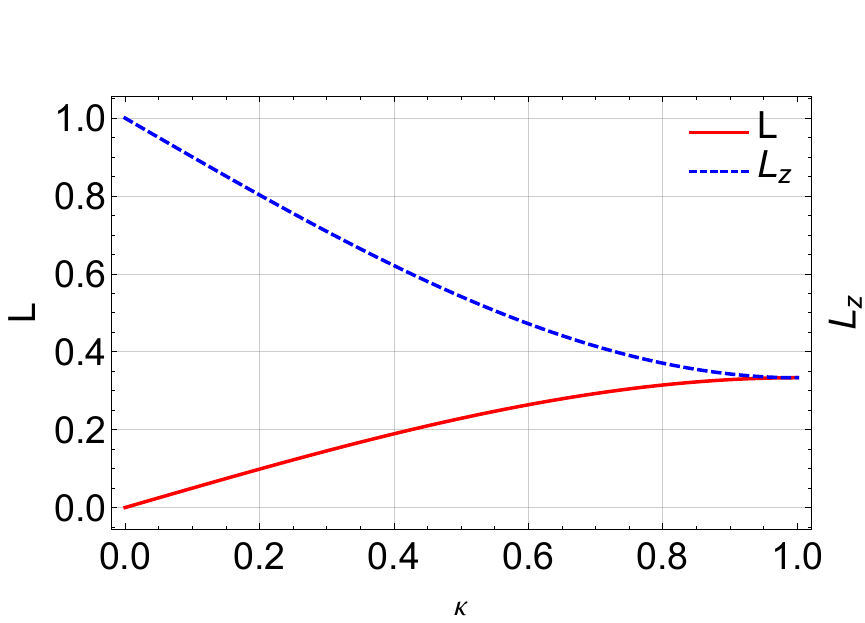} 
  \vspace{0mm}  \hfill
 \caption{\label{Fig3} $L$ and $L_z$ plotted  against  $\kappa\in(0,1)$ for the double-truncated unit sphere.
   }
\end{figure}

In Fig.~\ref{Fig3}, the depolarization factors $L$ and $L_z$ are plotted against $\kappa$.
As is the case for the truncated sphere presented in \S\ref{Truncated_sphere_sec}, for the double-truncated sphere
 $L \to 0$ and $L_z \to 1$ in the limit $\kappa \to 0 $, while
$L \to 1/3$ and $L_z \to 1/3$ in the limit $\kappa \to 1 $. Furthermore, the plots of $L$ and $L_z$ for the double-truncated sphere
in Fig.~\ref{Fig3} are similar to the corresponding plots in Fig.~\ref{Fig2} for the truncated sphere. Differences are
most obvious in the regime of small values of $\kappa$ wherein $L$ (resp. $L_z$) increases (resp. decreases) less rapidly
for the double-truncated sphere 
as $\kappa$ increases.

\section{Spheroidal geometry}

\subsection{Hemispheroid }

Consider the spheroid
\begin{equation}
\frac{x^2 + y^2}{\alpha^2} + \le z+ \kappa \ri^2 \leq 1,
\end{equation}
with $\alpha> 0$ and $0 \leq \kappa \leq 1/2$. 
The spheroid is oblate for $\alpha >1 $ and prolate for $\alpha < 1$.
Suppose that the spheroid is
bifurcated in two  halves by the symmetry plane $z=-\kappa$.
Then $V_\delta$ is    the  hemispheroid   bounded below by the plane $z=-\kappa$,
as  schematically illustrated in Fig.~\ref{Fig1d}.
 The 
 largest inscribed sphere is specified by Eq.~\r{largest_sphere}
with radius
\begin{equation}
\kappa = \left\{  \begin{array}{lcr}
1/2 &\mbox{for} & \alpha > 1 / \sqrt{2} \vspace{8pt}\\
\alpha \sqrt{1- \alpha^2} & \mbox{for} & \alpha \leq 1 / \sqrt{2}
\end{array}\right..
\end{equation}

Again, $\=L  =   L \left( \ux\ux+  \uy\uy\right)+ L_z\,\uz\uz$ with $L_z=1-2L$.
We found from  Eq.~(\ref{depol}) for $ \alpha \leq 1 / \sqrt{2}$ that
\begin{equation} \displaystyle{
L = \frac{1+ \nu + \alpha^2 \les \alpha \le \alpha + \sqrt{2- \alpha^2} \ri- \nu - 3 \ris
- \le\alpha \gamma \ri^2 \sqrt{2 - \alpha^2} 
\log \displaystyle{\frac{\le 1- \alpha^2 + \gamma \ri
\le 1- \alpha \ri}{\alpha \le \nu - \gamma \ri}}}{4 \nu \gamma^4},}
\end{equation}
where
\begin{equation}
\left.
\begin{array}{l}
 \nu =  \sqrt{2 - 3 \alpha^2 + \alpha^4}
 \vspace{6pt}\\
 \gamma = \sqrt{1-\alpha^2 }
\end{array}
\right\}
;
\end{equation}
and for $ \alpha > 1 / \sqrt{2}$ 
\begin{equation} \displaystyle{
L = \frac{3 \le 1+ \sqrt{1+4\alpha^2} \ri
+ \alpha^2 \le 10 - 8 \sqrt{1+4 \alpha^2} \ri - 8 \alpha^4}{4 \gamma^2  
\le    3 +  8 \alpha^2 - 16 \alpha^4 \ri }
+ \frac{\alpha^2 \log
\displaystyle{ \frac{1-2\alpha^2 + \gamma}
{\sqrt{1+ 3 \alpha^2 - 4 \alpha^4 }-1}}}{4 \gamma^3 }
}.
\end{equation}

\begin{figure}[!htb]
\centering
\includegraphics[width=8cm]{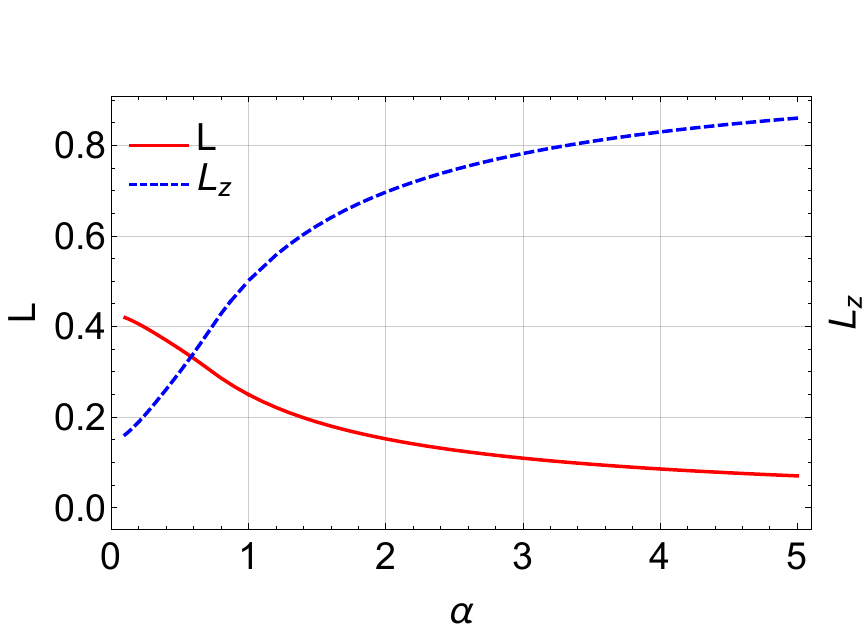} 
  \vspace{0mm}  \hfill
 \caption{\label{Fig4} $L$ and $L_z$ plotted against $\alpha$ for the hemispheroid.
   }
\end{figure}

Plots of the depolarization factors $L$ and $L_z$ versus $\alpha$ are displayed in Fig.~\ref{Fig4}.
As $\alpha$ increases,
the value of $L$ decreases monotonically  whereas the value of $L_z$ increases monotonically. For  $\alpha < 0.569$, $L > L_z$; and for $\alpha > 0.569$,  $L < L_z$.
In the limit $\alpha \to 0$ we find $L = \le 2 + \sqrt{2} \ri/8$ and $L_z = \le 2 - \sqrt{2} \ri/4$, while
in the limit $\alpha \to \infty$ we find $L=0$ and $L_z = 1$.
For $\alpha = 1$ the results for the hemisphere are recovered.

\subsection{Double-truncated spheroid}

Consider  the spheroid
\begin{equation}
\frac{x^2 + y^2}{\alpha^2} +  z^2 \leq 1,
\end{equation}
centered at the coordinate origin
with $\alpha> 0$. 
The spheroid is oblate for $\alpha >1 $ and prolate for $\alpha < 1$.
Suppose that the spheroid 
is trifurcated  symmetrically by the planes $z = \pm\kappa$, where  $0<\kappa<1$.
Then $V_\delta$ is 
  the double-truncated spheroid bounded by the planes $z = \pm \kappa$,  as  
 shown in Fig.~\ref{Fig1e}.   The 
 largest inscribed sphere is centered at the  origin of the coordinate system.
 
Equation (\ref{depol}) yields $\=L  =   L \left( \ux\ux+  \uy\uy\right)+ L_z\,\uz\uz$,
where
\begin{equation}
L = \frac{1-L_z}{2}=\frac{- \nu + \alpha^2 \tan^{-1} \nu}{2 \le\alpha^2 -1 \ri^{3/2}}
\end{equation}
with
\begin{equation}
\nu = \kappa \sqrt{\frac{\alpha^2 -1}{\alpha^2- \kappa^2 \le \alpha^2-1 \ri}};
\end{equation}
and again $L_z = 1- 2 L$. 

\begin{figure}[!htb]
\centering
\includegraphics[width=7cm]{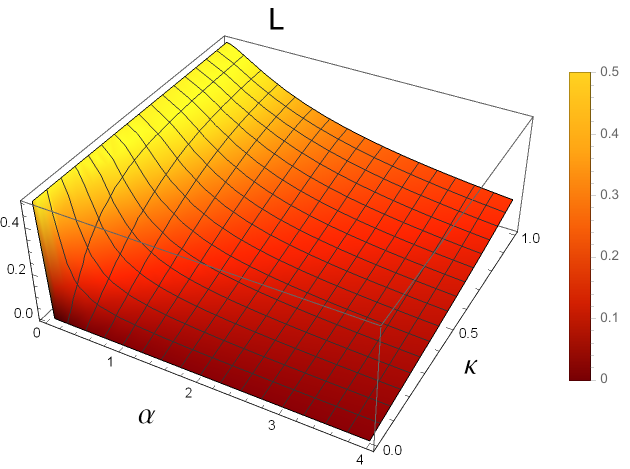} 
  \vspace{0mm}  \hfill
 \caption{\label{Fig5} $L$  plotted against $\alpha\in(0,4]$ and  $\kappa\in(0,1)$ for the double-truncated spheroid.
   }
\end{figure}

In Fig.~\ref{Fig5}, the depolarization factor
 $L$ is plotted against $\alpha$ and $\kappa$.
As $\alpha$ increases,  $L$ decreases monotonically and $L_z$ increases monotonically, 
for all values of $\kappa$. As $\kappa$ increases, $L$ increases monotonically
and $L_z$ decreases monotonically
 for all values of $\alpha>0$.
In the limit $\kappa \to 0$ we find $L = 0$ and $L_z = 1$, while
in the limit $\kappa \to 1$ we find
\begin{equation}
L = \frac{1}{2} \le \frac{1}{1-\alpha^2} + \alpha^2 \frac{\tan^{-1} \sqrt{\alpha^2 -1 }}
{\le \alpha^2 -1 \ri^{3/2}} \ri,
\end{equation}
in agreement with the standard results for spheroids \cite{Moroz}. In the limit $\alpha \to 0$, $L= 1/2$ and $L_z =0$; 
while $L = 0$ and $L_z =1$ in the limit $\alpha \to \infty$. And the results for the hemisphere are recovered  with $\alpha = 1$.

\section{Ellipsoidal geometry}

In the cases of truncated ellipsoidal geometries, 
closed-form expressions for $L_x$, $L_y$, and $L_z$ could not be obtained, but numerical methods were used to evaluate 
these  depolarization factors.

\subsection{Hemi-ellipsoid}

Consider the ellipsoid
\begin{equation}
\frac{x^2}{\alpha^2} + \frac{y^2}{\beta^2} + \le z+ \kappa \ri^2 \leq 1.
\end{equation}
with $\alpha>0$, $\beta> 0$, and $0 \leq \kappa \leq 1/2$.
Suppose that the ellipsoid is
 truncated by the symmetry plane $z=-\kappa$. We chose $V_\delta$ as
 the upper hemi-ellipsoid bounded below by the plane $z=-\kappa$.
 The 
 largest inscribed sphere is specified by Eq.~\r{largest_sphere}
with radius
\begin{equation}
\kappa = \left\{  \begin{array}{lcr}
1/2,  & \mbox{for}& \alpha > 1 / \sqrt{2}, \quad \beta > 1 / \sqrt{2} \vspace{8pt}\\
\alpha \sqrt{1- \alpha^2}& \mbox{for} & \alpha \leq 1 / \sqrt{2}, \quad \alpha < \beta
 \vspace{8pt}\\
\beta \sqrt{1- \beta^2}& \mbox{for} & \beta \leq 1 / \sqrt{2}, \quad \beta < \alpha
\end{array}\right.
.
\end{equation}

\begin{figure}[!htb]
\centering
\includegraphics[width=7cm]{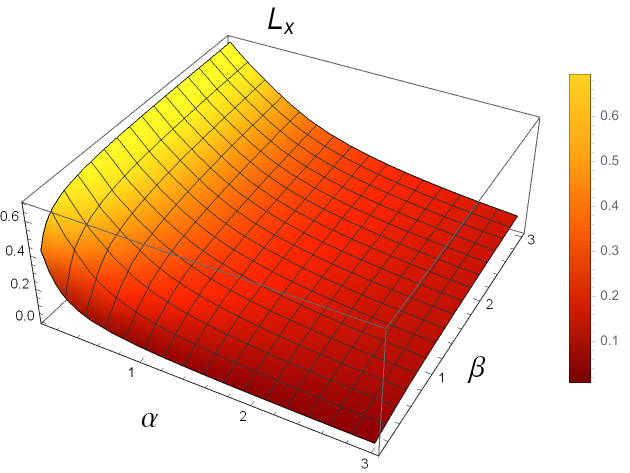} 
\hspace{10mm}
\includegraphics[width=7cm]{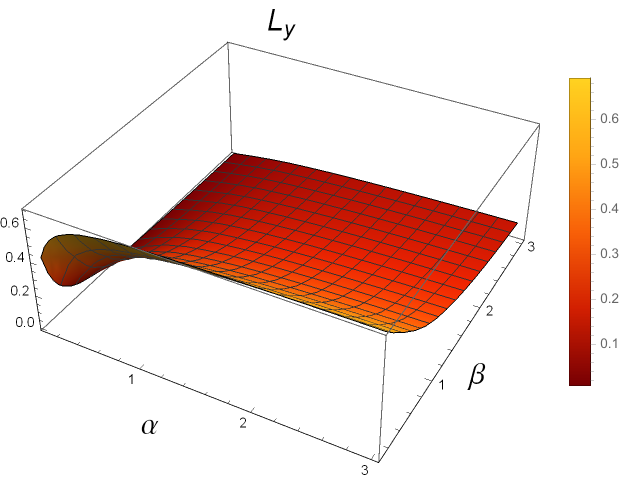} 
 \caption{\label{Fig6}
  $L_x$ and $L_y$ plotted against $\alpha\in(0,3]$ and $\beta\in(0,3]$
   for the hemi-ellipsoid.
   }
\end{figure}

In Fig.~\ref{Fig6},
plots of the depolarization factors $L_x$ and $L_y$ versus $\alpha$ and $\beta $ are displayed; the plot of $L_z$ may be inferred from $L_z = 1- L_x - L_y$.
All  depolarization factors vary smoothly as $\alpha$ and $\beta$ increase. In particular,
$L_x$ decreases markedly as $\alpha$ increases but is relatively insensitive to changes in $\beta$; 
$L_y$ decreases markedly as $\beta$ increases but is relatively insensitive to changes in $\alpha$; 
and $L_z$ generally increases as both $\alpha$ and $\beta$ increase.

\subsection{Double-truncated ellipsoid}

Consider  the ellipsoid
\begin{equation}
\frac{x^2}{\alpha^2} + \frac{y^2}{\beta} + z^2 \leq 1,
\end{equation}
centered at the  origin of the coordinate system 
with $\alpha> 0$ and $\beta>0$. 
Suppose that the ellipsoid
is trifurcated by the planes
 $z = \pm\kappa$, where  $0<\kappa<1$.
 Then $V_\delta$ is 
  the double-truncated ellipsoid bounded by the planes $z = \pm \kappa$,  and
  the largest inscribed sphere is centered at the  origin of the coordinate system.

\begin{figure}[!htb]
\centering

\includegraphics[width=7cm]{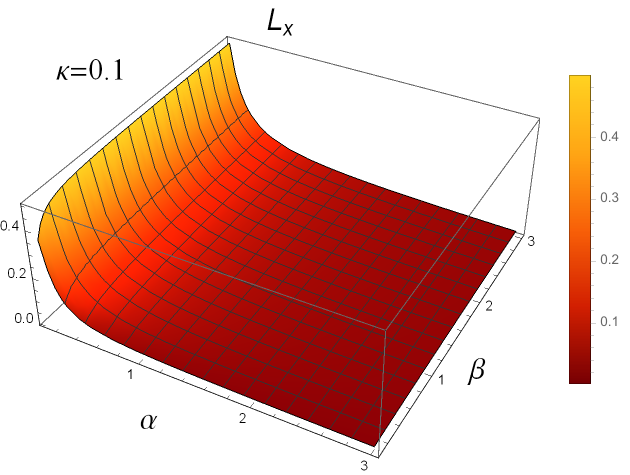}  \hspace{5mm}
 \includegraphics[width=7cm]{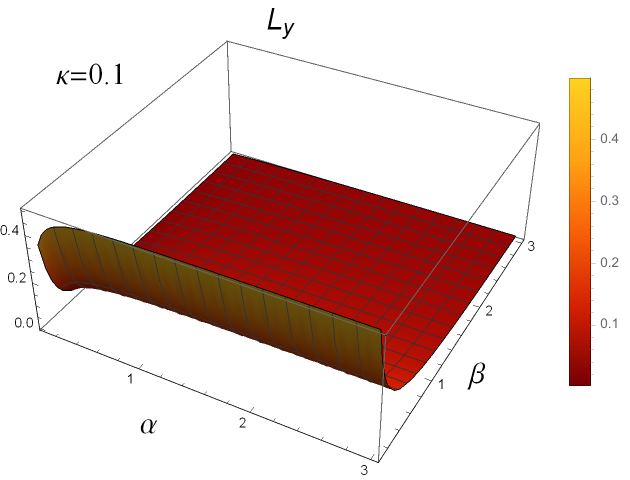}\\
\includegraphics[width=7cm]{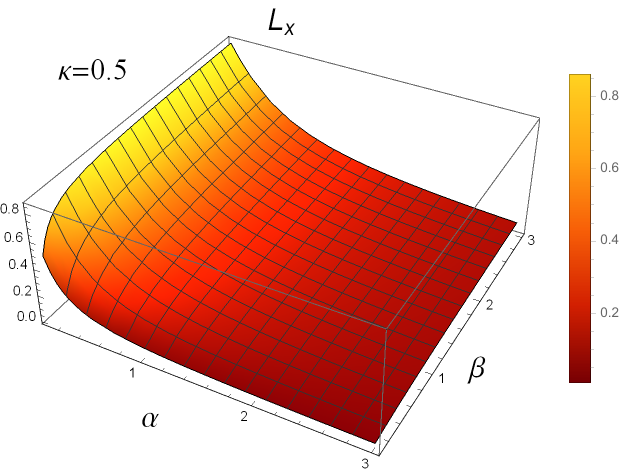} \hspace{5mm}
\includegraphics[width=7cm]{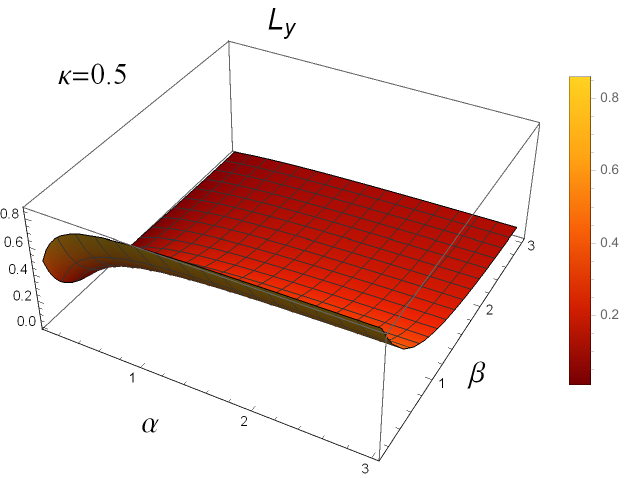} \\
\includegraphics[width=7cm]{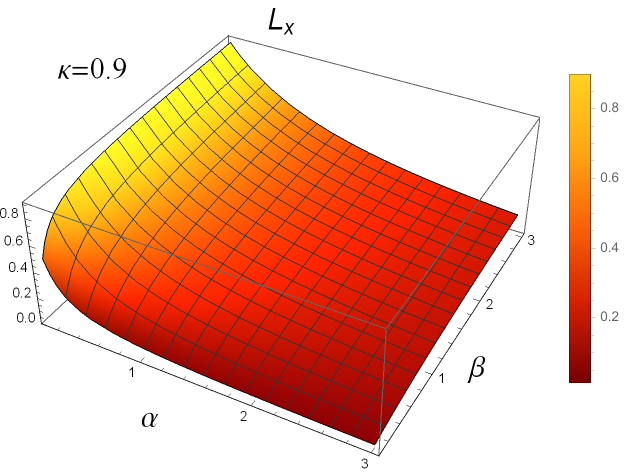}  \hspace{5mm}
 \includegraphics[width=7cm]{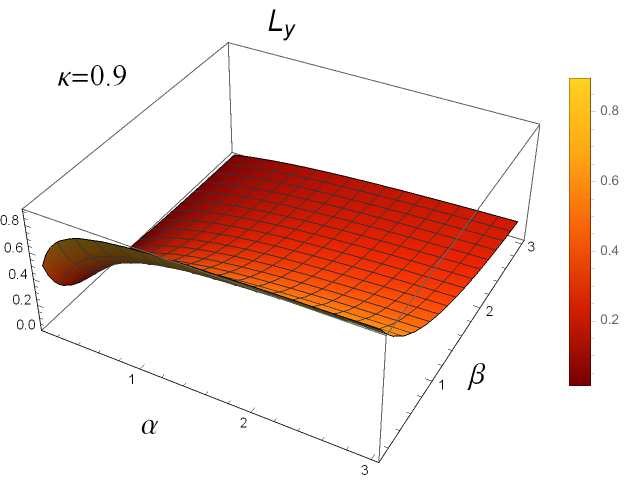}
 \caption{\label{Fig7}
 $L_x$ and $L_y$ plotted against  $\alpha\in(0,3]$ and $\beta\in(0,3]$  for the double-truncated ellipsoid with $\kappa \in \lec 0.1, 0.5, 0.9 \ric$.}
\end{figure}

For $\kappa \in \lec 0.1, 0.5, 0.9 \ric$, 
plots of $L_x$ and $L_y$ versus $\alpha$ and $\beta $ are provided in Fig.~\ref{Fig7}; the plot of $L_z$ may be inferred since $L_z = 1- L_x - L_y$.
The plots in Fig.~\ref{Fig7} for the double-truncated ellipsoid 
are qualitatively similar to those in Fig.~\ref{Fig6} for the hemi-ellipsoid.
The effects of varying $\kappa$ are most appreciable at low values of $\alpha$ for $L_x$, at low values
of $\beta$ for $L_y$, and at low values of both $\alpha$ and $\beta$ for $L_z$. Generally, the values of
the depolarization factors
 $L_x$, $L_y$, and $L_z$  are 
substantially  more sensitive to variations in $\alpha$ and $\beta$ than they are to variations in $\kappa$.

 \section{Closing remarks}
 
 Depolarization dyadics are central to theoretical studies in scattering and homogenization, but closed-form expressions for these entities have been available only for a few simple shapes. 
 Closed-form expressions for depolarization dyadics have been developed herein for truncated spheres and truncated spheroids, and the formalism has been extended to truncated ellipsoids; the evaluation of depolarization dyadics for this latter case requires numerical integration.
 These theoretical results will enable studies of scattering from complex-shaped small particles
 \cite{BH} as well as of homogenization
 of particulate composite materials containing complex-shaped particles \cite{MAEH}.
 These results are particular valuable in view of the ongoing rapid  development of nanocomposite materials for optical applications \cite{Pinar}.

\vspace{10mm}

 \noindent {\bf Acknowledgments:} 
TGM was supported  by
EPSRC (grant number EP/V046322/1).  AL   was supported by the US National Science Foundation (grant number DMS-1619901)  as well as   the Charles Godfrey Binder Endowment at Penn State.\\

\noindent{\bf Declaration of competing interest:}
The authors declare that they have no known competing financial interests or personal relationships that could have
appeared to influence the work reported in this paper.

\newpage

\newpage

\vspace{10mm}

\vspace{10mm}

\vspace{10mm}

\newpage

\vspace{10mm}

\newpage

\begin{thebibliography}{99}

\bibitem{Chen}
H.C. Chen, \emph{Theory of Electromagnetic Waves}.
New York, NY, USA: McGraw--Hill, 1983.


\bibitem{Tai}
C.T. Tai, \emph{Dyadic Green Functions in Electromagnetic Theory,
2nd Ed}.  Piscataway, NJ, USA: IEEE Press, 1994.


\bibitem{Faryad} 
M. Faryad and A. Lakhtakia,
\emph{Infinite-Space Dyadic Green
Functions in Electromagnetism}. San Rafael, CA, USA:  Morgan \& Claypool, 2018. 

\bibitem{VanBladelBook}
J. Van Bladel, \emph{Singular Electromagnetic Fields and Sources}.
 Oxford, UK: Oxford University Press, 1991 (reissued in association with
IEEE Press,
New York, NY, USA, 1995).

\bibitem{EAB}
T.G. Mackay and A. Lakhtakia,
 \emph{Electromagnetic Anisotropy and Bianisotropy: A Field Guide}, {2nd Ed}.
Singapore: World Scientific,  2019.

\bibitem{MAEH}
  T.G. Mackay and A. Lakhtakia,
 \emph{Modern  Analytical Electromagnetic  Homogenization with Mathematica}, \emph{2nd Ed}.
 Bristol, UK:
 IOP Publishing,  2020.

\bibitem{M97} B. Michel, ``A Fourier space approach to the pointwise
singularity of an anisotropic dielectric medium,'' \emph{Int. J. Appl.
Electromagn. Mech.}, vol.~ 8,  pp.~219--227, 1997.

\bibitem{Moroz}
A. Moroz,
``Depolarization field of spheroidal particles,'' \emph{J. Opt. Soc. Am. B}, vol.~26, pp.~517--527, 2009. 

\bibitem{Osborn}
J.A. Osborn,
``Demagnetizing factors of the general ellipsoid,''
\emph{Phys. Rev.}, vol.~67, pp.~351--357, 1945.

\bibitem{Stoner}
 E.C. Stoner, ``The demagnetizing factors for ellipsoids,'' \emph{Phil. Mag.}, vol.~36, pp.~803--821, 1945.

\bibitem{MW97} B. Michel and W.S.    Weiglhofer,
``Pointwise singularity of dyadic Green function in a general
bianisotropic medium,'' \emph{Arch. Elektr. \"Ubertrag.}, vol.~51, pp.~219--223, 1997.
 Corrections: vol.~52, p.~310, 1998.

\bibitem{W98}
W.S. Weiglhofer, ``Electromagnetic depolarization dyadics and
elliptic integrals,'' \emph{J. Phys. A: Math. Gen.}, vol.~31, pp.~7191--7196,  1998.

\bibitem{Lee}
S.W. Lee, J. Boersma, C.L. Law, and G.A. Deschamps,
``Singularity in Green's function and its numerical evaluation,''
\emph{IEEE Trans.  Antennas.  Propagat.},
vol.~28, pp.~311--317, 1980.

\bibitem{WM02}
W.S. Weiglhofer and T.G. Mackay, ``Needles and pillboxes in
anisotropic mediums,'' \emph{IEEE Trans. Antennas Propagat.}, vol.~50,
pp.~85--86, 2002.


\bibitem{L98}
A. Lakhtakia and N.S. Lakhtakia, ``A procedure for evaluating depolarization dyadics of polyhedra,''
\emph{Optik}, vol.~109, pp.~140--142, 1998.

\bibitem{Wang}
J.J.H. Wang, ``A unified and consistent view on the singularities of the electric dyadic Green's function in
the source region,'' \emph{IEEE Trans. Antennas Propagat.}, vol.~30,
  463--468, 1982.


\bibitem{Yaghjian}
A.D. Yaghjian,  ``Electric dyadic Green's function in the source region,'' \emph{Proc. IEEE}, vol.~68,  pp.~248--263, 1980.

\bibitem{Sihvola}
J. Avelin, A. N. Arslan, J. Br\"{a}nnback, M. Flykt, C. Icheln, J. Juntunen, K. K\"{a}rkk\"{a}inen, T. Niemi, 
O. Nieminen, T. Tares, C. Toma, T. Uusitupa, and A. Sihvola,
``Electric fields in the source region: the depolarization dyadic 
for a cubic cavity,'' \emph{Elec. Eng.}, vol.~81,  pp.~199--202, 1998.

\bibitem{BH}
C.F. Bohren and D.R. Huffman, \emph{Absorption and Scattering of
Light by Small Particles}. New York: Wiley, 1983,



\bibitem{Pinar}
M.P. Meng\"{u}\c{c} and M. Francoeur (Eds.), \emph{Light, Plasmonics and Particles}.  Amsterdam: Elsevier, 2023.



\end{thebibliography}
\end{document}